\documentclass[compsoc, conference, a4paper, 10pt, times]{IEEEtran}

\usepackage[utf8]{inputenc}
\usepackage[T1]{fontenc}
\usepackage{microtype}

\usepackage{amsmath,amssymb,amsfonts}
\usepackage{siunitx}
\usepackage{textcomp}
\usepackage{wasysym}
\usepackage{xspace}

\usepackage{xcolor}

\usepackage{graphicx}
\usepackage{svg}

\usepackage{float}
\usepackage{placeins}

\usepackage{booktabs}
\usepackage{multirow}
\usepackage{array}
\usepackage{tabularx}
\usepackage{colortbl}
\usepackage{makecell}
\usepackage{tabularray}

\usepackage{enumitem}
\usepackage{tcolorbox}
\tcbuselibrary{listings,breakable}

\usepackage{soul}
\usepackage{comment}
\usepackage{algorithmic}
\usepackage{subfiles}

\usepackage[font=small]{caption}
\usepackage{subcaption}
\usepackage{listings}

\lstdefinelanguage{yaml}{
  keywords={true,false,null,y,n},
  keywordstyle=\color{blue}\bfseries,
  basicstyle=\ttfamily\small,
  comment=[l]{\#},
  commentstyle=\color{gray}\ttfamily,
  stringstyle=\color{red}\ttfamily,
}

\lstdefinelanguage{SPL}{
        keywords=[1]{
        case, match, strftime, num, count
        },
        keywords=[2]{
            AND, and, OR, or, AS, As, as, by
        },
        keywords=[3]{
            timeformat
        },
        keywords=[4]{
            spath, search, dedup, stats, eval, convert, chart, rename
        },
        keywordstyle=\color{purple},
        keywordstyle=[2]\color{orange},
        keywordstyle=[3]\color{green!75!black},
        keywordstyle=[4]\color{blue},
        commentstyle=\color{green}, 
        stringstyle=\color{red}, 
        identifierstyle=\color{black}, 
        basicstyle=\footnotesize\sffamily, 
        columns=fullflexible,
        breaklines=true,
        numbers=left,
        numberstyle=\footnotesize\color{gray}, 
        xleftmargin=0.5cm, 
        framexleftmargin=0.5cm,
        backgroundcolor=\color{white} 
    }

\usepackage{cite}

\usepackage{hyperref}
\hypersetup{
  colorlinks=true,
  linkcolor={black!80!black},
  citecolor={red!70!black},
  urlcolor={blue!70!black},
  breaklinks=true
}
\usepackage{url}

\hypersetup{hidelinks}
\newcommand{\code}[1]{\texttt{\nolinkurl{#1}}}

\usepackage{cleveref}

\title{Peacock: UEFI Firmware Runtime Observability Layer for Detection and Response}

\author{
\IEEEauthorblockN{Hadar Cochavi Gorelik, Orel Fadlon, Denis Klimov, Oleg Brodt, Asaf Shabtai, Yuval Elovici}\vspace{0.5ex}
\IEEEauthorblockA{Ben Gurion University of the Negev}
}

\begin{document}

\maketitle

\begin{abstract}
Modern computing platforms rely on the Unified Extensible Firmware Interface (UEFI) to initialize hardware and coordinate the transition to the operating system.
Because this execution environment operates with high privileges and persists across reboots, it has increasingly become a target for advanced threats, including bootkits documented in real systems.
Existing protections, including Secure Boot and static signature verification, are insufficient against adversaries who exploit runtime behavior or manipulate firmware components after signature checks have completed.
In contrast to operating system (OS) environments, where mature tools provide dynamic inspection and incident response, the pre-OS stage lacks practical mechanisms for real-time visibility and threat detection.
We present Peacock, a modular framework that introduces integrity-assured monitoring and remote verification for the UEFI boot process.
Peacock consists of three components:
(i) a UEFI-based agent that records Boot and Runtime Service activity with cryptographic protection against tampering;
(ii) a cross-platform OS Agent that extracts the recorded measurements and produces a verifiable attestation bundle using hardware-backed guarantees from the platform’s trusted module; and
(iii) a Peacock Server that verifies attestation results and exports structured telemetry for enterprise detection.
Our evaluation shows that Peacock reliably detects multiple real-world UEFI bootkits, including Glupteba, BlackLotus, LoJax, and MosaicRegressor.
Taken together, these results indicate that Peacock provides practical visibility and verification capabilities within the firmware layer, addressing threats that bypass traditional OS-level security mechanisms.
\end{abstract}

\section{\label{sec:introduction}Introduction}

The Unified Extensible Firmware Interface (UEFI)~\cite{UEFI2024} succeeded the legacy Basic Input/Output System (BIOS) as the dominant firmware framework in modern computing~\cite{UEFI2023}.  
This shift emerged from the structural and functional constraints of the BIOS design, which relied on a 16-bit execution mode, a one-megabyte addressing limit, and a non-modular architecture.  
Introduced in the early 2000s, UEFI was engineered to provide a scalable and extensible interface that coordinates hardware initialization with operating system (OS) booting.  
Its widespread implementation now extends from inexpensive embedded boards such as Raspberry Pi devices to consumer laptops, desktops, and servers, underscoring its role as a universal component of contemporary computer platforms~\cite{sentinelone_uefi_dumping}.  

UEFI serves as a foundational element in modern computer architecture and therefore represents a critical point in the system’s security model.  
Operating immediately above the hardware layer, UEFI operates with elevated privileges, maintains persistence across system restarts, and coordinates the complete boot sequence before transferring control to the OS.  
A compromise within this layer can enable adversaries to evade kernel-level and hypervisor-based protections, thereby endangering the integrity of the entire system.  

Boot-level security incorporates several layers of protection intended to establish trust from the earliest stages of execution.  
The UEFI Forum~\cite{web:uefi_forum}, supported by efforts from industry and the research community \cite{TPM2, eclypsium_secure_boot_dbx, CapsuleUpdate}, standardized a collection of mechanisms designed to secure this process.  
Among the most central of these is UEFI Secure Boot~\cite{web:secure_boot_microsoft, web:secure_boot_redhat}, which enforces code integrity by verifying the digital signatures of boot components against a repository of trusted certificates.  
Complementing this, UEFI Capsule Updates \cite{CapsuleUpdate} enable the delivery of firmware updates whose authenticity and integrity are protected through cryptographic validation.  
Security assurances are further strengthened by hardware-based elements such as Trusted Platform Modules (TPMs)~\cite{web:intel_tpm,TPM2} and embedded controllers (ECs)~\cite{web:intel_me}, both of which contribute to establishing a hardware root of trust for attestation and secure key management.  
Collectively, these capabilities form the foundation of a Trusted Computing Base (TCB)~\cite{TCB} that aims to protect the firmware domain against unauthorized modification and tampering.  

Despite these measures, UEFI security effectiveness depends heavily on implementation and configuration. Secure Boot, for example, is often disabled or misconfigured, either to support legacy software or due to usability concerns. 
Even when properly enabled, attackers can exploit its policies and trusted keys. 
The BlackLotus bootkit bypassed Secure Boot by leveraging a vulnerable yet signed bootloader~\cite{web:eset_blacklotus}, while LOGOFAIL~\cite{web:kaspersky_logofail} and PIXIFAIL~\cite{web:eclypsium_pixifail} exploited flaws in UEFI code itself to achieve arbitrary code execution (ACE) or remote code execution (RCE). Although UEFI Capsule Updates employ cryptographic verification, they are not immune to supply-chain compromise.  
The ShadowHammer campaign demonstrated that attackers could distribute malicious firmware through a compromised update infrastructure despite the use of cryptographic validation~\cite{shadowhammer}.

Recent attack campaigns illustrate the progression of UEFI threats beyond static vulnerabilities. MoonBounce demonstrated persistence by residing in Serial Peripheral Interface (SPI) flash and redirecting execution flow during runtime to hook boot services~\cite{MoonBounce_2022}. 
CosmicStrand injected malware into the OS by altering runtime service structures~\cite{CosmicStrand_2022}, while Glupteba disabled security controls by patching multiple boot components~\cite{Glupteba_2024}. 
These cases highlight that bypassing or abusing Secure Boot is not the only path to compromise; attackers increasingly target the runtime environment itself.

Yet, aside from Secure Boot’s signature checks, no robust security controls currently provide runtime monitoring or behavioral inspection during the UEFI phase. 
As long as malicious code passes or bypasses signature verification, it can execute without meaningful oversight in the pre-OS environment. 
This gap stands in contrast to OS-level defenses, where mature mechanisms for runtime visibility, logging, and threat detection are standard practice.

In this paper, we present Peacock, a firmware focused detection and attestation framework that brings structured visibility to the UEFI execution environment.
Peacock combines integrity aware logging during the pre boot phase with post boot attestation and remote verification, allowing security systems to observe and validate the behavior of firmware components.
By linking low level UEFI telemetry with higher level security operations, Peacock provides a practical path for detecting unauthorized modifications, persistence mechanisms and anomalous behavior within the early boot sequence.

\noindent Our main contributions are as follows: 
\begin{itemize}[topsep=0pt,noitemsep,leftmargin=*]
\item \textbf{Firmware-level integrity monitoring.}
We designed a UEFI-based mechanism that records Boot and Runtime Service activity and links these measurements to hardware-backed integrity protections, ensuring that firmware behavior can be validated after the OS has started.
\item \textbf{End-to-end attestation workflow for pre-boot execution.}
We developed a post-boot attestation process that collects firmware-level evidence, generates verifiable attestation artifacts, and validates the authenticity and freshness of the recorded measurements using TPM-backed cryptographic guarantees.
\item \textbf{Remote verification and secure ingestion of firmware telemetry.}
We implemented a trusted verifier that checks the integrity and authenticity of the collected evidence, reconstructs expected measurements, and forwarding firmware events to a dedicated server for further investigation and detection.  
This design prevents unverified or potentially tampered data from entering enterprise monitoring pipelines.
\item \textbf{Evaluation on representative, real-world UEFI threats.}
We evaluated the framework against widely analyzed real-world UEFI bootkits, demonstrating its ability to support detection of service table manipulation, Secure Boot bypasses, filesystem-based persistence, and multi-component implants.

\end{itemize}
\section{\label{sec:background}Background}

\subsection{The UEFI}
The Unified Extensible Firmware Interface (UEFI), standardized by the UEFI Forum \cite{UEFIAbout}, defines the architecture and functionality of modern platform firmware. 
It replaced the legacy BIOS, which lacked the flexibility and scalability needed for contemporary hardware and operating systems.  
UEFI introduced a modular and extensible design that standardizes the pre-boot environment, managing hardware initialization and configuration before the operating system is launched.  
This framework provides several advantages over legacy BIOS, including faster boot sequences, support for large storage devices, improved abstraction between hardware and software, and integrated security features.  

UEFI is often described as a “mini operating system” because it performs a wide range of low-level tasks that prepare the platform for the operating system.  
It initializes and manages hardware components, loads device drivers, performs disk operations, establishes network connections, provides a command-line shell for user interaction, and executes UEFI applications, all within the pre-boot environment.  

UEFI firmware is stored as a binary image within one or more Serial Peripheral Interface (SPI) flash chips located on the system motherboard.  
This non-volatile storage preserves the firmware across power cycles and supports updates via UEFI Capsule Updates \cite{CapsuleUpdate}.
However, this persistence also creates a significant security concern: once the firmware image is compromised, an attacker can maintain control even after reinstalling the operating system or replacing storage devices.  

The UEFI boot process progresses through several phases, each performing a dedicated role in initializing the platform.
The Security (SEC) phase begins execution, verifying firmware integrity and establishing a temporary execution environment.  
The Pre-EFI Initialization (PEI) phase then discovers and configures system memory.  
Next, the Driver Execution Environment (DXE) phase, a core stage of UEFI operation, loads and executes drivers and protocols needed to configure the hardware and prepare the platform for the OS transition.  
As this phase plays a critical role in both system configuration and attack surface exposure, we discuss it further in \autoref{sec:background:dxe}.  
Following DXE, the Boot Device Selection (BDS) phase determines and loads the OS bootloader.  
An optional Transient System Load (TSL) phase may occur, where components such as the UEFI shell or bootloaders finalize the system setup before transferring control to the OS.  
Finally, the Runtime (RT) phase maintains UEFI runtime services that persist after boot, allowing ongoing interaction between the firmware and the OS.

\subsection{Driver Execution Environment (DXE)}
\label{sec:background:dxe}
The DXE phase is a critical component of the UEFI boot process, acting as a bridge between the PEI phase and the BDS phase.  
The DXE phase plays an important role in system initialization and prepares the runtime environment for loading the operating system.  

After the PEI phase establishes a minimal hardware environment and provides essential boot services, such as memory and CPU initialization, the DXE phase is responsible for initializing the remaining hardware components using dedicated DXE drivers.  
DXE drivers are modular pieces of code designed to manage specific hardware devices or system services.  
These drivers are essential for platform functionality, as they initialize and configure system hardware components for boot and subsequent use by the operating system.  
Such drivers are typically embedded within the UEFI image in firmware volumes or within the peripheral device itself, and are served to the UEFI as an Optional ROM (OpROM).  

The core activity of the DXE phase is the loading and execution of DXE drivers, though they are not the only images processed at this stage.  
In general, UEFI images fall into three categories: \textit{Boot Service Drivers}, which operate only during the boot process; \textit{Runtime Drivers}, which remain active after the OS takes control; and \textit{UEFI Applications}, which execute on demand.  

Each driver image in the UEFI environment is uniquely identified by a 128-bit Globally Unique Identifier (GUID).  
GUIDs serve as universal references that distinguish software components, interfaces, and drivers across different platforms.  
They provide a reliable mechanism for tracking, mapping, and managing resources during firmware initialization.  
When the image represents a UEFI application, it can alternatively be associated with the file path from which it was loaded (for example, \textit{\textbackslash EFI\textbackslash Boot\textbackslash BootX64.efi}), indicating its location within the EFI System Partition (ESP) or another storage medium.  

Beyond hardware initialization, various system services critical for subsequent boot stages are established during the DXE phase, including memory management, I/O protocols, security services, and event management.  
By setting up these services, the DXE phase ensures that the system is in a stable state before the operating system is selected and loaded during the BDS phase.  
The BDS phase relies on the environment configured by the DXE phase to select and load the operating system.  

UEFI images can originate from several sources.  
They may be executed manually from the UEFI shell, which allows users to launch applications or scripts; retrieved from SPI flash, where core firmware and DXE drivers reside; dynamically loaded from peripheral devices through OPROMs; or read from the EFI System Partition (ESP), a dedicated disk partition that stores bootloaders and related configuration files necessary for system startup.  

Given its pivotal role in hardware and system services initialization, the DXE phase is a focal point for security.  
Ensuring that only authenticated and authorized DXE drivers are executed is critical to maintaining the integrity of the boot process.  
Mechanisms such as Secure Boot, which verifies the digital signatures of DXE drivers, are employed to prevent the execution of malicious code during this phase.

\subsection{UEFI Services Tables}

The UEFI specification includes the UEFI System Table (also referred to as \textit{EFI\_SYSTEM\_TABLE}), arguably the most important data structure in UEFI, which abstracts hardware complexities and provides a standardized interface for firmware and operating system interaction.
This abstraction offers several benefits, such as hardware independence, extensibility, enhanced security, and improved boot performance ~\cite{zimmer2017beyond}.
UEFI firmware operates as an early-stage execution environment, and its services function as the firmware equivalent of system calls that operating systems provide to applications.\footnote{The term system call conventionally refers to OS-specific interfaces and does not technically apply to UEFI. We use this analogy to aid readers with OS backgrounds who are new to UEFI.}

\begin{figure}[ht]
   \includegraphics[scale=1.4]{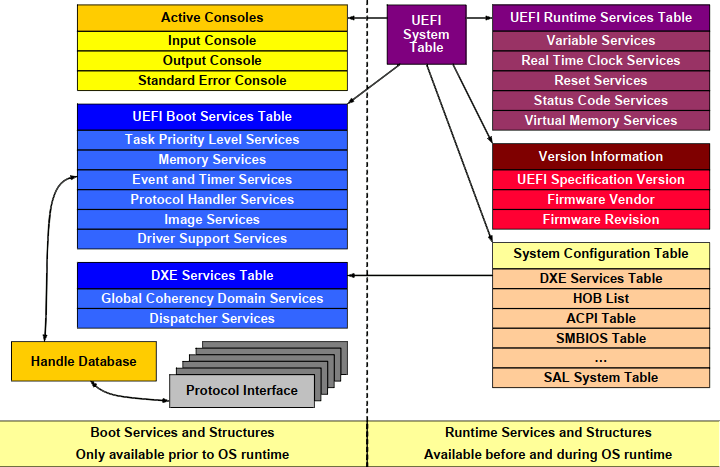}
\captionsetup{font=small, width=.9\columnwidth}
\caption{UEFI system table and related components \cite{uefi_pi_2023}.}
\label{fig:uefitables}
\end{figure}

During the DXE phase, DXE drivers load UEFI services into memory to enable boot completion and OS support.
The UEFI System Table holds pointers to two main service tables: the Boot Services (also referred to as \textit{EFI\_BOOT\_SERVICES}) and the Runtime Services (also referred to as \textit{EFI\_RUNTIME\_SERVICES}).
Boot services provide firmware interfaces available exclusively during platform initialization, whereas runtime services maintain availability after OS handoff.
These services are organized into tables according to service category, as illustrated in \autoref{fig:uefitables}.
UEFI firmware exposes these tables as callable interfaces throughout the boot process and, for Runtime Services, to the OS after boot completion.

\vspace{1.5ex}
\noindent\textbf{EFI Boot Services Table.}
The EFI Boot Services Table provides access to functions required for boot operations, encompassing hardware initialization and OS handoff.
By convention, this table is referred to as \texttt{gBS}.
Each UEFI specification revision defines a fixed number of services within this table.
The UEFI 2.10 specification defines 45 Boot Services distributed across several categories, as presented in Table \ref{tab:UEFI_services}.
Key service categories include memory management, protocol management, event services, and timer services.
Boot Services remain available exclusively during boot phases.

\vspace{1.5ex}
\noindent\textbf{EFI Runtime Services Table.}
The services provided by this table remain available after the boot process completes, exposing low-level services for OS use during runtime.
By convention, this table is referred to as \texttt{gRT}.
The UEFI 2.10 specification defines 20 Runtime Services organized into functional groups, as shown in Table \ref{tab:UEFI_services}.
Key runtime services include variable services, time services, and system reset services.

\begin{table}[H]
    \centering
        
    \setlength{\tabcolsep}{4pt} 
    
    \resizebox{\columnwidth}{!}{

    \begin{tabular}{@{} l c l c @{}}
        \toprule
        \multicolumn{2}{c}{\textbf{Boot Services}} & \multicolumn{2}{c}{\textbf{Runtime Services}} \\ 
        \toprule
        Name & Count & Name & Count \\ 
        \cmidrule(r){1-2} \cmidrule(l){3-4}
        Event, Timer, and Task Priority Services & 9 & Variable Services & 8 \\ 
        Memory Allocation Services & 5 & Time Services & 4 \\ 
        Protocol Handler Services & 18 & Virtual Memory Services & 2 \\ 
        Image Services & 6 & Miscellaneous Runtime Services & 6 \\ 
        Miscellaneous Boot Services & 7 & & \\ 
        \midrule
        \textbf{Total} & 45 & & 20 \\        
        \bottomrule
    \end{tabular}
    }
  
%\captionsetup{font=small, width=.9\columnwidth}
    \caption{UEFI services by groups}
     \label{tab:UEFI_services}
\end{table}

\subsection{Legitimate Function-Pointer Hooking}% in Enterprise}
\begin{figure}[ht]
\includegraphics[scale=0.28]{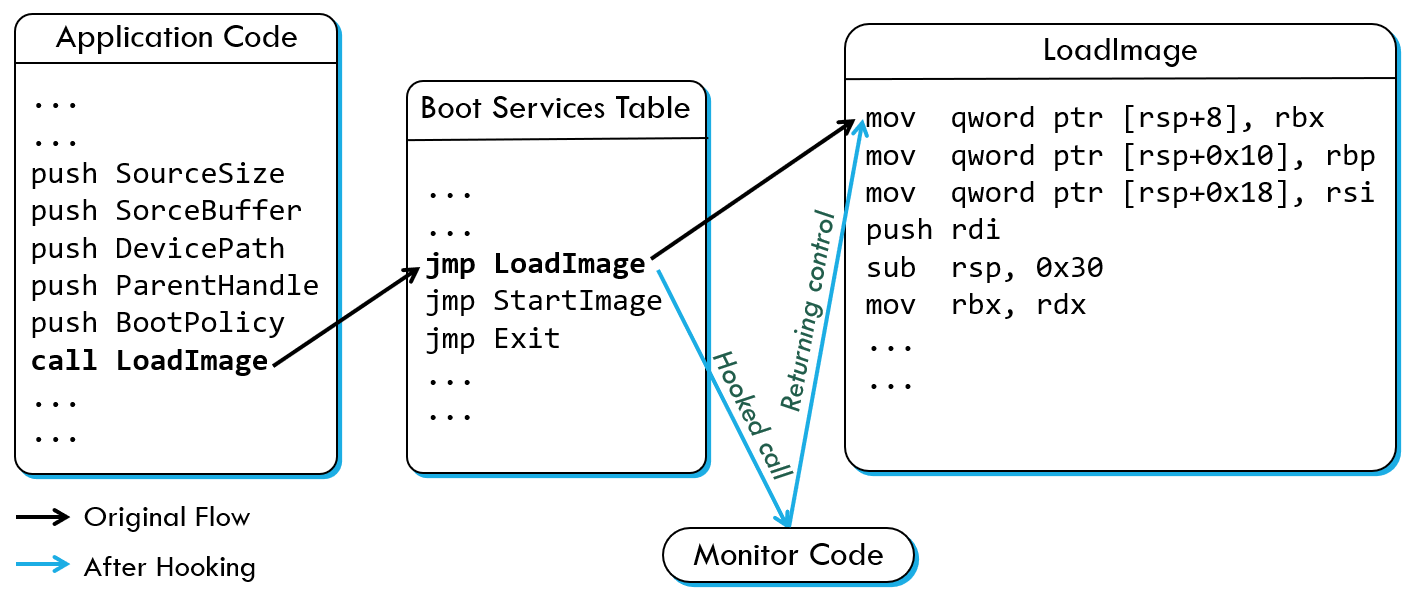}
\captionsetup{font=small, width=0.9\columnwidth}
\caption{Function pointer hooking.}
\label{fig:tradHooking}
\end{figure}

\label{sec:legit_hooking}
Function pointer hooking operates by overriding the addresses stored in service or API function pointers to redirect execution flow to an alternate implementation.
As Illustrated in \autoref{fig:tradHooking}, this technique replaces a function pointer value with the address of code controlled by the hook implementer so that subsequent calls are redirected to the new target.
When the caller invokes the original function, control is transferred to the hooked code, which may perform monitoring, mediation, or modification before optionally invoking the original implementation.
Although function pointer hooking is commonly associated with attacker techniques used to achieve persistence or concealment, it is also widely adopted by legitimate defensive and instrumentation tools for monitoring, protection, and runtime extension.
Instrumentation and monitoring libraries such as Microsoft Detours illustrate legitimate uses of function interception to observe and extend program behavior for debugging, telemetry, and security applications \cite{web:Detours}.

Dynamic instrumentation frameworks used by security researchers and defenders, for example Frida, provide similar interception capabilities to inspect runtime behavior, trace sensitive API usage, and implement inline mitigations during testing and incident response \cite{web:frida}.
In production enterprise environments, Endpoint Detection and Response (EDR) products commonly rely on controlled hooking techniques to observe sensitive system and API calls, implement behavioral controls, and detect suspicious activity at runtime \cite{arfeen2021endpoint}.

Because hooking can be abused by attackers, defensive products must carefully manage hook provenance and integrity, typically by maintaining authoritative allow-lists of expected hooks and by protecting or restoring legitimate hook targets when possible \cite{Glupteba_2024, blacklotus, MoonBounce_2022,EfiGuard}.

\subsection{Trusted Platform Module (TPM)}
The Trusted Platform Module (TPM) is a long-standing hardware security component deployed widely across modern computing platforms \cite{web:tcg}.
It is standardized and maintained by the Trusted Computing Group (TCG), which defines both the TPM architecture and its role within trusted computing ecosystems \cite{web:tcg}.
Recent surveys describe the maturity and adoption of TPM 2.0 in contemporary systems \cite{pirker2023peek}.

\textbf{TPM Device Identity:} A TPM incorporates several capabilities that support device identity and integrity reporting.
Each TPM includes a unique Endorsement Key (EK), along with a manufacturer-issued EK certificate that allows external verifiers to confirm that the device implements a genuine, standards-compliant TPM \cite{web:tcg}.
This certificate may be stored directly on the TPM or retrieved via an external manufacturer service.

\textbf{PCR Extension:} TPMs expose Platform Configuration Registers (PCRs), which store integrity measurements using an append-only hashing operation known as PCR extension \cite{web:tcg_pc}.
When new data is extended into a PCR, the TPM computes a new digest by hashing the previous PCR value together with the new input, producing a cumulative measurement state that cannot be reset or overwritten arbitrarily.
Different PCR indices are assigned for different classes of measurements according to TCG and OS-specific conventions \cite{web:tcg_pc}.

\textbf{TPM Quote:} A TPM can attest to its internal integrity state.
Through a quote operation, the TPM signs selected PCR values using an attestation key, and incorporates a verifier-supplied nonce to ensure freshness.
This signed report enables an external system to verify both the authenticity of the TPM and the integrity state encapsulated in the quoted PCR set.

\subsubsection{Remote Attestation}
The TPM's unique identity, derived from its Endorsement Key and corresponding certificate, combined with its functionality, enables remote attestation of device state \cite{chen2024bind}.
In a basic attestation protocol, a verifier first receives the target device's EK certificate and verifies its authenticity.
The verifier then encrypts a secret using the public Endorsement Key, ensuring that only the TPM associated with that certificate can decrypt it.
Next, the verifier requests a quote of the PCR state that incorporates the decrypted secret, verifies the signature on the quote, and compares the reported PCR values to a known-good reference state.
However, the TPM cannot use the Endorsement Key to sign attestation quotes.
Instead, the device's TPM must generate a separate Attestation Key (AK) for this purpose.
The verifier must then verify that this Attestation Key genuinely belongs to the target TPM.

\subsection{Security Information and Event Management}

Security Information and Event Management (SIEM) systems provide centralized collection, correlation, and analysis of security events from diverse sources across enterprise infrastructure \cite{tariq2023open}.
SIEM platforms aggregate logs from network devices, endpoint security tools, applications, and infrastructure components, enabling security teams to identify threats through cross-source correlation and behavioral analysis .
Modern SIEM systems support real-time event processing, rule-based detection, and integration with automated response mechanisms.

Enterprise security operations rely on SIEM platforms to correlate events that may appear benign in isolation but indicate coordinated attacks when analyzed collectively \cite{tariq2023open}.
By maintaining historical baselines and performing statistical analysis across device fleets, SIEM systems can identify anomalous patterns that deviate from established organizational norms.
Common SIEM platforms include Splunk \cite{subramanian2020introducing}, and IBM QRadar \cite{chakrabarty2021securing}, which provide standardized ingestion APIs and query languages for processing heterogeneous log formats.
\section{\label{sec:threat_model}Threat Model}

\textbf{Attacker Capabilities.} We consider an adversary capable of executing code within the UEFI firmware environment during the DXE phase or at later pre-OS stages.
The attacker is assumed to operate within the constraints of the UEFI execution model and may interact with firmware components using standard Boot Services and Runtime Services.
Such an adversary may attempt to alter control flow, introduce unauthorized components, or perform operations that affect the integrity of the boot process.
We intentionally define the attacker’s capabilities at a behavioral level rather than enumerating specific techniques, since firmware threats continually evolve and may combine or extend known methods.

This model captures the class of attacks demonstrated by real-world UEFI bootkits, including service table manipulation, unauthorized image loading, persistent state modification, and filesystem-based pre-OS activity.
Prominent exemplars such as Glupteba, BlackLotus, LoJax, and MosaicRegressor fall within this category, but the model is not restricted to these specific families.

\textbf{Trust Assumptions.}
We assume that the Log Collection Module is correctly embedded into the UEFI firmware image and is configured to load before any third-party DXE components.
This ensures that monitoring begins prior to the execution of potentially malicious drivers.
The Log Collection Module itself is trusted to operate as intended, and the system includes mechanisms for detecting attempts to interfere with its instrumentation.

We assume that the OS Agent runs immediately after the OS is initialized.
Its role is limited to retrieving the recorded logs and producing a complete attestation bundle; it performs no security-critical verification.
Finally, we trust the Peacock Server to act as the authoritative verifier, validating the attestation evidence and confirming log integrity before forwarding structured telemetry to the SIEM for rule-based detection and alerting.
\section{\label{sec:methodology}Peacock: System Design and Architecture}

\begin{figure*}[!htb]
\centering
    \includegraphics[width=1\textwidth]{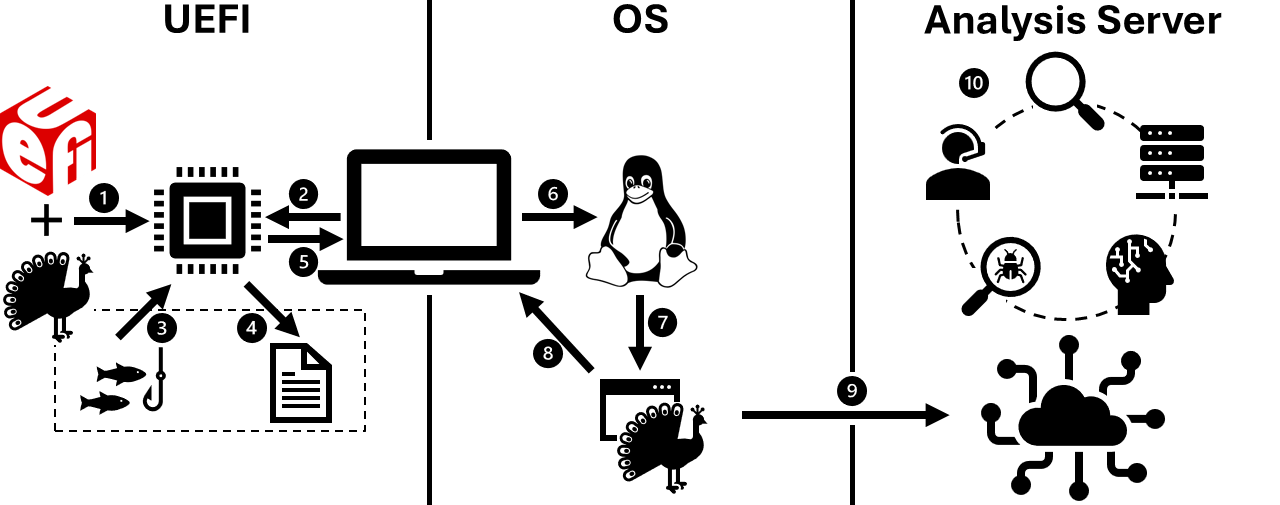}
    \caption{Peacock framework architecture.}
    \label{fig:Peacock_architecture}
\end{figure*}

This section presents the design and implementation of \textit{Peacock}, a modular framework that extends detection and response capabilities into the UEFI pre-boot environment.
The framework is designed to address the critical gap in firmware-level security monitoring by providing comprehensive visibility, tamper-evident logging, and automated threat detection during the vulnerable pre-OS phase when existing firmware security mechanisms prove insufficient against runtime attacks.

The Peacock framework comprises three integrated modules: (i) a \textbf{UEFI Agent} implemented as a UEFI DXE driver that act as log collection module and performs real-time service monitoring with cryptographic integrity assurance; (ii) an \textbf{OS Agent} that forwards firmware logs immediately upon OS startup; (iii) a \textbf{Central Analysis Server} that provides threat detection and enterprise-scale log ingestion through SIEM integration.
This multi-phase approach is intended to ensure comprehensive coverage of firmware activities while maintaining compatibility with existing security infrastructure and enabling scalable deployment in enterprise environments.

Figure~\ref{fig:Peacock_architecture} presents the overall architecture and data flow of the \textbf{Peacock framework}, which operates across three primary stages: the UEFI environment, the operating system (OS), and the Analysis Server.
(1) The process begins by embedding the Peacock UEFI Agent into the UEFI image, which is then installed on the target system.
(2) Upon system startup, the UEFI firmware is executed, and once it reaches the DXE phase, (3) the Peacock driver is loaded and begins to hook all UEFI services and allow normal firmware execution to proceed.
(4) When subsequent software components invoke UEFI services, these calls are intercepted and logged by the agent.
(5) Each log entry is measured by the TPM and extended into dedicated PCR, and the complete log is written to selected location on the ESP.
(6) After the UEFI firmware completes execution, control is handed over to the selected bootloader, and the OS begins to load.
(7) Once initialized, the preinstalled Peacock OS Agent is launched, (8) retrieves the UEFI logs, and requests the TPM to generate a quote over the PCR that was extended during boot.
(9) The OS Agent then encapsulates the logs and the TPM attestation data into single package, securely transmitting it to the Peacock Analysis Server.
(10) On the server side, the received data undergoes centralized validation, integrity verification, and advanced threat correlation analysis to detect potential firmware or boot-level anomalies across enterprise endpoints.

\subsection{Peacock UEFI Agent}
\label{sec:log_collection_module}
Our Peacock UEFI agent solution is implemented as DXE driver for the Tianocore EDK II \cite{edk2} ecosystem.
The DXE phase is the modular stage of the UEFI boot process in which firmware drivers initialize hardware and system services. Most UEFI code executes during this phase, whose flexible, driver-based design enables extensive extensibility and platform customization.
EDK II, the de-facto standard of UEFI specification implementation, provides the flexibility to compile drivers for various target platform architectures (IA32, IPF, X64, EBC, ARM, or AArch64), making the approach platform-agnostic.
The main goal of the UEFI agent is to serve as a log collection module, monitor and audit firmware execution to logs that can be used later for analysis.
Our hooks integrity is validated against a whitelist to detect unauthorized service table modifications, the collected logs are protected with cryptographic integrity assurance powered by the platform TPM through hash chaining.

\textbf{Phase I - Embedding:} We integrate our DXE driver into the targeted UEFI firmware; \textbf{Phase II - Collecting:} by hooking the UEFI services our agent monitors and logs UEFI service calls during firmware execution; \textbf{Phase III - Exporting:} logs are written to the EFI System Partition (ESP) and when OS starts, our OS Agent sends them to the analysis server.

\subsubsection{Phase I - Embedding}
\label{sec:embedding}
To embedded our Peacock UEFI agent DXE driver into the target UEFI firmware, we used publicly available tools such as UEFITool~\cite{web:git_uefi_tool} and dedicated embedding utilities.
To maximize monitoring coverage, the driver must be loaded as early as possible in the DXE phase. Therefore, we configured it to execute immediately after the DXE Core is initialized.
Once the firmware image is prepared, it must be deployed to the target system.
This process is straightforward in virtual environments but requires additional steps in physical environments, such as software-based capsule updates or hardware-based flashing with an SPI programmer.
OEMs (e.g., Lenovo, Dell, HP) and UEFI firmware vendors can pre-install the module within their firmware as part of the standard distribution process, simplifying deployment in enterprise environments.
The Peacock UEFI agent provides configurable monitoring capabilities to accommodate diverse deployment requirements. This configurability balances monitoring comprehensiveness against performance and storage constraints based on specific threat models and operational requirements.

\subsubsection{Phase II - Collection}
\label{sec:collection}
During system startup, the Peacock UEFI agent is loaded into the UEFI memory and start monitoring the system.
extracts firmware metadata including (i) version information, (ii) vendor identity, (iii) release date, and more. After that, the agent hooks the system main services, and by that starts monitoring UEFI firmware activities.

\paragraph{\textbf{Service Hooking}}
Upon initialization, the agent hooks all services in the Boot Services and Runtime Services tables.
By hooking these tables, our agent can inspect every function call made to the hooked services.
To implement this, the Peacock UEFI agent: (i) queries the system for pointers to the \texttt{EFI\_BOOT\_SERVICES} and \texttt{EFI\_RUNTIME\_SERVICES} tables; and (ii) replaces original service pointers in the tables with pointers to its own interception proxy functions.
Each service hook comprises four elements:
\begin{itemize}[topsep=0pt,noitemsep,leftmargin=*]
    \item Pointer holding original service address.
    \item Our interception proxy function.
    \item Hook installation function.
    \item Hook uninstallation function for cleanup operations.
\end{itemize}
When a software component invokes a UEFI service, our agent intercept the call, and the proxy function executes instead. 
Our function logs the call details and forwards the call to the original service without any modification.
This mechanism provides comprehensive visibility into firmware runtime behavior without disrupting normal UEFI operations.

\paragraph{\textbf{Service Call Logging}}
Each monitored service invocation is recorded with a unique log identifier (LID), high-resolution timestamp (T), and call correlation identifier (CID) for event tracking.
The log entry captures a triplet containing: (i) caller identification (driver GUID or file path, and memory range), (ii) service name and entry parameters, and (iii) service exit parameters including return status.
Representative raw log entries generated during execution are shown below.

\textbf{Caller Identification:}
{\fontsize{8}{10}\selectfont
\begin{verbatim}
(LID:257) (T:3892988953) (CID:144) [CheckCaller]
Caller GUID - 'F80697E9-7FD6-4665-8646-88E33EF71DFC',
start address 7EF78000, end address 7EF7DCC0
\end{verbatim}
}

\textbf{Service Call Entry and Exit:}
{\fontsize{8}{10}\selectfont
\begin{verbatim}
(LID:267) (T:3897972787) (CID:147) Enter 
LocateProtocol -
Service Address:'7F6AEB0E',
Protocol:'94AB2F58-1438-4EF1-9152-18941A3A0E68',
Registration:'0', Interface:'7FE77C60'

(LID:268) (T:3898212105) (CID:147) Exit 
LocateProtocol -
Service Address:'7F6AEB0E', Interface:'0',
RetStatus:'Not Found'
\end{verbatim}
}
Each entry includes the LID, timestamp, CID, caller metadata, service address, argument values, and return status.  
The CID allows the framework to correlate the “Enter” and “Exit” records of a single invocation, enabling accurate reconstruction of firmware-level behavior during boot.

\paragraph{\textbf{Log Integrity and Tamper Detection}}
To ensure the integrity and authenticity of audited UEFI logs, we utilized the platform Trusted Platform Module (TPM), as a hardware-based root of trust.
Each audited log entry is hashed and used to extend a selected TPM’s Platform Configuration Registers (PCRs), creating a cryptographic chain that reflects the chronological sequence of logged events.
For each subsequent log entry, a hash value is computed as \(PCR_{new}=H(PCR_{old}||m)\) where \(PCR_{new}\) is the new PCR value, \(PCR_{old}\) is the previous PCR value, \(m\) is the hash of the log being measured, and \(H\) is a cryptographic hash function.
This process provides forward integrity: any alteration, deletion, or reordering of previous log entries would result in a mismatch with the final PCR value.
Upon completion of the logging process, the final PCR state can be attested by the TPM using its Attestation Key (AK).
The signed attestation data, together with the corresponding log records, represent the system’s execution state with TPM-backed integrity.
The verifier can recompute the log hashes and compare the derived PCR value with the attested measurement to confirm the integrity and authenticity of the logs. This mechanism provides a hardware-backed guarantee that the UEFI logs have not been tampered with and that they originate from a genuine, trusted platform.

\paragraph{\textbf{Hook Integrity}}
The Peacock UEFI agent validates hook integrity to distinguish between legitimate and potentially malicious service table modifications.
Some legitimate drivers modify the UEFI service tables for functional purposes, while malicious firmware may attempt similar modifications for evasion or attack.

The agent maintains a configurable whitelist of drivers authorized to perform service table hooking.
When the agent detects external service table modifications, it performs the following validation:
\begin{enumerate}[topsep=0pt,noitemsep,leftmargin=*]
    \item \textbf{Driver Identification}: Identifies the hooking driver.
    \item \textbf{Whitelist Verification}: Check the identified driver against the predefined whitelist.
    \item \textbf{Remediation Action}: If whitelisted, re-establish monitoring hooks. If not whitelisted, generate a security alert.
    \item \textbf{Policy Enforcement}: Depending on organizational policy, either halt boot immediately (fail-secure mode) or continue boot while logging the alert (fail-open mode).
\end{enumerate}

This approach accommodates legitimate firmware behaviors while detecting unauthorized modifications that may indicate compromise.

\subsubsection{Phase III - Export}
\label{sec:export}
Upon completion of the UEFI firmware boot process, the Peacock UEFI agent stores the generated log in the EFI System Partition (ESP).
The log is written in a structured format that the OS Agent later collects as part
of the attestation bundle.

\subsection{Peacock OS Agent}
\label{sec:os_agent}
The Peacock OS Agent is implemented as a cross-platform solution supporting both Windows and Linux environments and is initialized immediately upon operating system startup.
Its primary function is to aggregate all data produced during the firmware boot process and encapsulate it into a structured package for transmission to the Peacock server.
This package contains: (i) the logs collected by the Peacock UEFI agent; and (ii) the TPM PCR measurements signed by the AK.
The package is securely sent to the Peacock Server, which validates the
attestation evidence, parses the firmware log, and forwards the structured
records to the SIEM for further investigation and threat detection.

\subsection{Peacock Analysis Server}
\label{sec:central_analysis}
The Analysis Server functions as the central verification and analytics component within the Peacock architecture. It operates at an enterprise scale to perform threat detection, correlation analysis, and generation of security insights across endpoints and device fleets.
The Peacock server securely stores and indexes the logs received from the OS Agent for long-term retention and analysis. This historical log repository serves as a trusted source of truth, enabling security experts to conduct targeted forensic investigations or to retrospectively identify prior instances of newly discovered attack patterns.

\subsubsection{Log Retrieval and Integrity Validation}
\label{sec:log_retrieval}
The server listens for incoming connections, and its exposed API enables endpoints. such as the Peacock OS Agent, to transmit collected logs for further aggregation and analysis.
Upon receiving the attestation package, the server verifies its authenticity by checking the TPM quote signature with the device’s pinned Attestation Key (AK) public key. 
This step confirms that the evidence was produced by the legitimate TPM associated with that device.
Following authentication, the server reconstructs the log hash chain to derive the expected PCR value and compares it with the attested one to identify any evidence of tampering or integrity violations.
Any discrepancy indicates of potential tampering, log corruption, or incomplete data collection, triggering immediate security alerts.

\subsubsection{Structured Parsing and Data Enrichment}
\label{sec:parsing_enrichment}
Upon successful validation, the server transforms raw firmware log entries into structured, machine-readable events.
Each parsed event record contains standardized fields as detailed in Table~\ref{tab:parsed_log_structure}:
\begin{itemize}[topsep=0pt,noitemsep,leftmargin=*]
    \item \textbf{Temporal Information}: UEFI timestamp and session identifier for chronological correlation
    \item \textbf{Service Details}: Service name and service address
    \item \textbf{Caller Context}: Driver GUID or path, memory address boundaries, and integrity hash when available
    \item \textbf{Hook Analysis}: Hook status, hooking driver identity, and whitelist validation results
    \item \textbf{Execution Context}: Service arguments, return status codes, and call correlation identifiers
    \item \textbf{Quality Assurance}: Original log entry and unique log identifiers for audit trails
\end{itemize}
This structured format enables efficient downstream processing while preserving forensic information.

\begin{table}[t]
\centering
\caption{Parsed Log Entry Structure and Field Descriptions}
\label{tab:parsed_log_structure}

\begin{tabular}{p{0.28\linewidth} p{0.65\linewidth}}
\hline
\textbf{Field Name} & \textbf{Description} \\
%Field Name & Description \\
\hline
\code{original_log} & Raw log entry as captured during UEFI execution before parsing \\
\code{uefi_timestamp} & High-precision timestamp indicating when the service call occurred \\
\code{event_type} & Service name \\
\code{caller} & Driver identifier (GUID) or file path of the entity that invoked the UEFI service \\
\code{caller_start_address} & Memory start address of the calling driver's loaded image \\
\code{caller_end_address} & Memory end address of the calling driver's loaded image \\
\code{hooked_service} & Boolean indicator of whether the service table entry has been modified \\
\code{hooked_by_driver} & Identifier of the driver responsible for service table modification \\
\code{whitelisted_hooking_driver} & Whether a hooking driver appears on the authorized whitelist \\
\code{status} & Return status of the service call execution (e.g., Success, Failure, NOT\_FOUND) \\
\code{args} & Input parameters passed to the UEFI service \\
\code{service_address} & Memory address of the service function at the time of call execution \\
\code{session_id} & Unique identifier for the specific boot session \\
\code{log_id} & Unique sequential identifier for this specific log entry \\
\code{call_id} & Correlation identifier linking multiple log entries for a single service call \\
\end{tabular}

\end{table}

\subsubsection{Threat Detection}
Threat detection in the server begins with attestation verification and continues through SIEM-based rule evaluation. The Peacock Server receives attestation bundles from multiple devices, validates the TPM quote, and confirms log integrity before any further analysis is allowed. This verification step functions as the first layer of detection, ensuring that only authentic and untampered firmware telemetry is processed.

Once attestation succeeds, the server parses the firmware log into a normalized, device-attributed structure suitable for integration with security monitoring workflows. The validated telemetry is then made available to downstream security analytics components, which may include SIEM platforms or other enterprise detection engines.

These analytics components evaluate the attested firmware events using rule-driven logic tailored for pre-boot activity, enabling the identification of behaviors such as:
\begin{itemize}[topsep=0pt,noitemsep,leftmargin=*]
    \item \textbf{Unauthorized Service Modifications}: Non-whitelisted changes to Boot or Runtime Service function pointers
    \item \textbf{Unexpected Driver Origins}: DXE modules loaded from atypical or untrusted ESP paths
    \item \textbf{NVRAM Manipulation}: Access to or creation of sensitive variables known to be used by persistent bootkits
    \item \textbf{Protocol and Filesystem Anomalies}: Access patterns that deviate from standard firmware behavior
\end{itemize}

In this architecture, the Peacock Server acts as the integrity-enforcing bridge between early-boot telemetry and enterprise detection infrastructure, enabling both real-time alerting and long term retention within existing security pipelines.
\section{\label{sec:eval}Evaluation}
The threat landscape described in \autoref{sec:threat_model} encompasses a range of documented UEFI bootkits and proof-of-concept implementations. 
These threats employ techniques such as service table manipulation, unauthorized image execution, filesystem operations, and persistent implantation through NVRAM or storage.

We evaluate Peacock's detection capabilities using four representative bootkit families: Glupteba \cite{Glupteba_2024}, BlackLotus \cite{blacklotus}, LoJax \cite{LoJax}, and MosaicRegressor \cite{MosaicRegressor}.  
These attacks represent diverse operational strategies and implementation approaches observed in real-world firmware threats.  

\subsection{Experimental Setup}
\label{sec:experimental_setup}
We evaluated Peacock across both virtualized and physical environments to assess detection effectiveness and practical feasibility.  

\subsubsection{Virtualized Environment}
Our primary evaluation platform used TianoCore EDK II \cite{edk2} firmware within QEMU \cite{qemu} virtual machines.  
These VMs emulated full system initialization from UEFI firmware through Windows 11 boot, which enabled controlled execution of bootkit samples and systematic variation of firmware configurations.  

\subsubsection{Physical Hardware Validation}
To demonstrate real-world applicability, we deployed the UEFI Agent on a physical System76 Adder WS (addw4) platform running Ubuntu 22.04 LTS.  
The system was equipped with a 14th Gen Intel Core i9-14900HX processor, 8 GB NVIDIA GeForce RTX 4060 GPU, 32 GB DDR5 RAM, a 1 TB PCIe4 M.2 SSD, and Coreboot BIOS version 4E3ADE8.  
This setup allowed us to confirm that the logging mechanism operates correctly on non-virtualized hardware.  

\subsubsection{Peacock Server and SIEM Configuration}
For remote attestation and centralized verification, we deployed the Peacock Server on a dedicated virtual machine running Ubuntu 22.04.4 LTS.  
The VM was configured with 8 virtual CPU cores and 16 GB of RAM.  
The OS Agent on the evaluated system communicated with the Peacock Server over an authenticated HTTPS channel.
For every boot session, the agent sent an attestation bundle that included: the raw firmware log, the Attestation Key (AK) public key and metadata, the TPM quote, the PCR value referenced by the quote, the nonce provided to the TPM, and the derived PCR digest.  
No verification was performed on the client side.  

Upon receiving a bundle, the Peacock Server acted as an attestation verifier.  
It reconstructed the AK public key, validated the TPM quote signature, checked the nonce for freshness, and confirmed that the PCR digest was consistent with the supplied PCR value.  
The server then recomputed the expected PCR value by hashing the raw boot logs using the same procedure applied by the UEFI Agent during boot and compared this recomputed value with the TPM-reported PCR value.  
Only if both the quote verification and the PCR comparison succeeded was the boot session marked as attested.  

For attested sessions, the Peacock Server invoked a dedicated parser that transformed the raw log into a structured JSON representation.  
The resulting JSON file was written to a directory monitored by the Splunk Universal Forwarder \cite{web:splunk}.
The Universal Forwarder transmitted the parsed logs to a remote Splunk instance, where SPL-based detection rules were executed, alerts were generated, and results were made available for investigation.  
If attestation failed at any point, the Peacock Server emitted only a minimal attestation failure record to Splunk and did not forward the full log for further analysis.  

\subsection{Detection Results}
\label{sec:detection_results}

This section evaluates the framework's ability to detect four representative UEFI bootkit families that exemplify distinct classes of firmware-level attacks.  
These families were selected because they collectively span major behavioral patterns observed in real-world UEFI malware, including service table manipulation, Secure Boot bypass techniques, filesystem-based persistence, and multi-component implants that rely on NVRAM markers.  
The framework detected all four families using the pipeline described in \autoref{sec:methodology}: firmware-level logging by the UEFI Agent, integrity validation and attestation through the OS Agent and Peacock Server, investigation and threat detection.

\subsubsection*{Threat Categories and Rationale for Selection}

\begin{itemize}[topsep=0pt,leftmargin=*]

\item \textbf{Bootloader Hijacking and Secure Boot Bypass.}
Attacks that compromise the early boot chain by abusing signed components or replacing trusted bootloaders.
BlackLotus exemplifies this category through its exploitation of vulnerable Windows boot binaries and unauthorized certificate enrollment.

\item \textbf{UEFI Service Table Hooking.}
Threats that manipulate Boot or Runtime Service function pointers to gain control over firmware execution.
Glupteba represents this class through its unauthorized modification of the \texttt{LoadImage} entry.

\item \textbf{Filesystem-Based Persistence During Boot.}
Malware that relies on pre-OS filesystem access to install or maintain persistent payloads.
LoJax is a representative example, using READY\_TO\_BOOT callbacks to perform NTFS-based file drops.

\item \textbf{Multi-Component Implants with NVRAM Markers.}
Framework-style threats that deploy several coordinated DXE components and rely on custom NVRAM variables as infection flags.
MosaicRegressor demonstrates this pattern through its “fTA” variable and clustering of related GUIDs.

\end{itemize}

\subsubsection*{Detection Rules}
The detection rules presented in the subsections below constitute a representative subset of the full rule corpus used during evaluation.
They were selected because they provide clear, high-confidence examples that illustrate how Peacock identifies distinct behavioral signatures associated with each bootkit family.
The complete rule set includes additional structural, behavioral, and temporal rules that reinforce detection accuracy but are omitted here due to space constraints.

\subsubsection{Glupteba Detection (Service Table Hooking)}
\label{sec:glupteba_detection}
Glupteba is a UEFI bootkit derived from the publicly available EfiGuard bootkit.
It deploys two components to the ESP, a DXE driver (\texttt{EfiGuardDxe.efi}) and a loader (\texttt{Loader.efi}), which together replace the legitimate Windows Boot Manager and ensure early execution.

The malware modifies the Boot Services Table by installing an unauthorized hook on \texttt{LoadImage}.
Although EfiGuard supports hooking multiple services, the Glupteba variant used here modifies only \texttt{LoadImage}.
The driver recalculates the table’s CRC32 field to mask these changes.

\textbf{Splunk-Based Detection}
Splunk’s rule evaluation engine generated multiple alerts that matched Glupteba’s behavior profile.
Alerts included the detection of unauthorized modification of the \texttt{LoadImage} function pointer by a non-whitelisted component, as well as the execution of \texttt{EfiGuardDxe.efi} from the ESP, a location inconsistent with legitimate DXE drivers.
These example alerts, drawn from the full ruleset evaluated in Splunk, formed a coherent and high-confidence identification of the Glupteba bootkit.

\begin{lstlisting}[language=SPL,
  caption={SPL detection rule identifying unauthorized service table modification by a DXE driver loaded from the EFI System Partition.},
  label={lst:glupteba-spl},
  frame=single,
  float,
  floatplacement=H
]
# Service Table Hooking from ESP-Loaded Driver
hooked_service=true hooked_by_driver="\\EFI*" whitelisted_hooking_driver=false
| stats count by hooked_service, hooked_by_driver, event_type, caller, uefi_timestamp, log_id, call_id, session_id
| sort -count
\end{lstlisting}

Listing~\ref{lst:glupteba-spl} illustrates an SPL rule that detects service table hooking by a driver loaded from the ESP.
The corresponding detection output generated by this rule is shown in Table~\ref{tab:esp_result}.

\begin{table*}[h]
    \centering
        
    \setlength{\tabcolsep}{6pt} 
    
    \begin{tabular}{@{} l c c c @{}}
        \toprule
        \textbf{caller} & \textbf{caller\_start\_address} & \textbf{caller\_end\_address} & \textbf{whitelisted\_hooking\_driver} \\ 
        \toprule

        \texttt{\textbackslash EFI\textbackslash Boot\textbackslash bootx64\_orig.efi} 
            & 0x10000000 & 0x101C8000 & false \\

        \texttt{\textbackslash EFI\textbackslash Boot\textbackslash EfiGuardDxe.efi} 
            & 0x7F62B000 & 0x7F67D000 & false \\

        \texttt{\textbackslash EFI\textbackslash BOOT\textbackslash BOOTX64.EFI} 
            & 0x7ABF6000 & 0x7AC01000 & false \\

        \bottomrule
    \end{tabular}
%    }
  
    \caption{Splunk results showing detection of an ESP-loaded DXE driver performing unauthorized service table modification.}
    \label{tab:esp_result}
\end{table*}

\subsubsection{BlackLotus Detection (Bootloader Hijacking \& Secure Boot Bypass)}
\label{sec:blacklotus_detection}

BlackLotus is a Secure Boot–bypassing bootkit that exploits CVE-2022-21894 to inject malicious components early in the boot chain.
The malware deploys vulnerable but signed Windows boot binaries and registers a malicious certificate in the Machine Owner Key (MOK) list, enabling execution of a Linux GRUB loader on a Windows system.
As part of its deployment, BlackLotus creates a nonstandard directory on the ESP and places its malicious binaries and configuration files within it, for example \texttt{ESP:\textbackslash system32\textbackslash}. Such directories are never used by legitimate Windows boot processes.

\textbf{Splunk-Based Detection}
Splunk’s rule evaluation engine produced several distinct alerts characteristic of BlackLotus.
Detection alerts included execution of \texttt{grubx64.efi} within a Windows environment, detection of Windows boot components loaded from ESP subdirectories such as \texttt{ESP:\textbackslash system32\textbackslash}, and unauthorized modification of the \texttt{MokList} NVRAM variable.
Additional matches, including detection of \texttt{VbsPolicyDisable}, aligned with the known BlackLotus execution sequence.
These representative alerts illustrate the broader set of rule matches used to reliably identify the threat.
A representative rule that flags loading of \texttt{grubx64.efi} on Windows systems is shown in Listing~\ref{lst:blacklotus-spl}.

\begin{lstlisting}[language=SPL,
  caption={Detection rule flagging LoadImage or StartImage events involving grubx64.efi in a Windows boot sequence.},
  label={lst:blacklotus-spl},
  frame=single,
  float,
  floatplacement=H
]
# BlackLotus Bootkit - Linux GRUB Bootloader on Windows
(event_type="LoadImage" OR event_type="StartImage") args="*grubx64.efi*"
| stats count by caller, args, status, uefi_timestamp, event_type, log_id, call_id, session_id
| sort -count
\end{lstlisting}

\subsubsection{LoJax Detection (Filesystem-Based Persistence During Boot)}
\label{sec:lojax_detection}

LoJax is the first publicly documented in-the-wild UEFI rootkit.
It maintains persistence by registering a READY\_TO\_BOOT callback, loading an NTFS driver, scanning the filesystem for Windows installations, and dropping malicious files prior to OS handoff.

\textbf{Splunk-Based Detection}
Splunk generated multiple alerts matching the known LoJax behavioral pattern.
The alerts included registration of a READY\_TO\_BOOT callback by a non-OEM component, followed by rapid \texttt{LocateHandleBuffer} queries for DiskIo and BlockIo protocols, and subsequent \texttt{OpenProtocol} calls targeting multiple partitions.
These alerts exemplify the broader ruleset that evaluated protocol enumeration patterns and filesystem access behavior, enabling reliable identification of LoJax.
Listing~\ref{lst:lojax-spl} presents a rule that detects READY\_TO\_BOOT callback registration by non whitelisted components.

\begin{lstlisting}[language=SPL,
  caption={Rule detecting registration of a READY\_TO\_BOOT event callback by a non-whitelisted component.},
  label={lst:lojax-spl},
  frame=single,
  float,
  floatplacement=H
]
# READY_TO_BOOT Event Callback Registration from Unknown Component
event_type="CreateEventEx" args="*7CE88FB3-4BD7-4679-87A8-A8D8DEE50D2B*" whitelisted_rtb_callback=false
| stats count by caller, args, uefi_timestamp, log_id, call_id, session_id
| sort -count
\end{lstlisting}

\subsubsection{MosaicRegressor Detection (Multi-Component Implant with NVRAM Marker)}
\label{sec:mosaicregressor_detection}

MosaicRegressor is a modular cyber espionage framework that uses a combination of UEFI applications and DXE drivers to maintain long term persistence in target environments.  
The firmware implant includes an NTFS capable component for low level filesystem access, a module that records infection state using a short named NVRAM variable (\texttt{fTA}), and DXE drivers that register READY\_TO\_BOOT callbacks to trigger additional logic late in the boot sequence.  
At this stage, a dedicated component locates the Windows installation, checks for specific marker files, and writes a user mode payload into the operating system startup path so that it is executed on every boot.

\textbf{Splunk-Based Detection}
Splunk’s detection rules produced several high-confidence alerts associated with MosaicRegressor’s multi-component design.
Representative alerts included unauthorized creation of the short-named \texttt{fTA} variable, clustering of related GUIDs linked to malicious DXE components, and READY\_TO\_BOOT callback registration by nonstandard modules.
Additional alerts, such as \texttt{HandleProtocol} operations involving \texttt{EFI\_LOADED\_IMAGE\_PROTOCO\_GUID}, contributed to a distinct signature.
These representative detections, drawn from the larger ruleset, collectively signaled the presence of MosaicRegressor.
A high confidence rule that detects the creation or access of the \texttt{fTA} infection marker variable is shown in Listing~\ref{lst:mosaicregressor-spl}.

\begin{lstlisting}[language=SPL,
  caption={Detection of MosaicRegressor infection marker showing 'fTA' NVRAM variable creation through unauthorized \texttt{SetVariable} operations.},
  label={lst:mosaicregressor-spl},
  frame=single,
  float,
  floatplacement=H
]
# MosaicRegressor Infection Marker - fTA NVRAM Variable
(event_type="GetVariable" OR event_type="SetVariable") args="*VariableName:'fTA'*"
| stats count by caller, event_type, args, status, uefi_timestamp, log_id, call_id, session_id
| sort -count
\end{lstlisting}

\subsection{Performance Evaluation}
\label{sec:performance_evaluation}
We assessed the operational impact of integrating the Peacock UEFI Agent on both virtualized and physical platforms, while the OS Agent and attestation workflow were evaluated within the virtualized environment.
Across all evaluated environments, system behavior remained consistent with baseline boot performance, and no observable changes in boot flow or system responsiveness were detected.
Because firmware execution characteristics vary across OEMs, chipsets, and bootloader implementations, performance overhead may differ on hardware configurations beyond those examined. Peacock’s logging design, described in Section~\ref{sec:embedding}, provides configurable monitoring depth, enabling deployments to balance visibility against performance and storage constraints according to platform requirements.
Overall, our observations indicate that the framework integrates smoothly into the boot process under the tested conditions, while a broader performance characterization across diverse hardware remains an important direction for future work.
\section{\label{sec:related}Related Work}

\subsection{UEFI Firmware Security Approaches}

Efforts to strengthen UEFI firmware security span several complementary directions, ranging from reverse engineering to vulnerability discovery and structural analysis.  
General-purpose analysis platforms such as IDA Pro \cite{IdaPro} and Ghidra \cite{Ghidra}, together with UEFI-focused extensions like efiXplorer \cite{efixplorer} and efiseek \cite{efiseek}, provide the foundation for examining firmware binaries and identifying potentially dangerous logic.  
Industry-oriented tools, including SentinelOne's Brick \cite{Brick} and Binarly's FWHunt \cite{FWHunt}, apply signature- and pattern-based scanning to detect known weaknesses and supply-chain threats in large firmware codebases.
Complementary parsing and extraction frameworks such as Chipsec \cite{chipsec}, UEFITool \cite{uefitool}, and Binwalk \cite{binwalk} facilitate structural inspection, region decoding, and validation of the complex layout of modern firmware images.

A substantial line of research has focused on automated vulnerability exploration.  
Fuzzing-based approaches, including AFL \cite{AFL}, TSFFS \cite{TSFFSUEFI}, Efi\_Fuzz \cite{efi_fuzz}, HBFA \cite{HBFA}, and Excite \cite{Excite}, have demonstrated success in discovering memory safety issues and logic flaws within UEFI implementations such as EDK II \cite{edk2}.  
Further academic work extends this direction with more specialized techniques: SPENDER \cite{yin2022finding} and RSFUZZER \cite{yin2023rsfuzzer} concentrate on System Management Mode (SMM) vulnerabilities, STASE \cite{shafiuzzaman2024stase} integrates static analysis with symbolic execution, and efiMemGuard \cite{lu2025automated} targets memory-related defects in DXE drivers.

Beyond traditional static analysis, recent research has begun exploring dynamic and hybrid inspection of firmware behavior.  
Notably, the UEFI Memory Analysis framework introduced in \cite{segal2025uefi} proposes fine-grained recovery of memory remnants left by firmware modules during initialization, enabling post-boot reconstruction of internal execution state.  
This body of work highlights the growing recognition that firmware runtime behavior contains valuable forensic and security-relevant information.

Despite these advances, prior approaches primarily focus on offline, emulated, or post-execution analysis.  
They do not incorporate continuous monitoring during the boot sequence nor provide mechanisms for detecting live malicious activity as firmware services execute.  
In contrast, Peacock introduces a runtime-oriented perspective: it records firmware events as they occur, protects these measurements using hardware-backed integrity, and integrates them into an attested, remotely verifiable detection workflow.  
This design shifts from static examination toward proactive identification of anomalous behavior within the pre-OS environment.

\subsection{Firmware Visibility and Runtime Monitoring}

Research on firmware visibility has examined methods for capturing system state before the operating system becomes active.  
Early works focused on memory acquisition during boot \cite{osborne2013memory, vidas2007acquisition, latzo2019universal}.  
UEberForensIcs, for example, embedded a DXE driver into UEFI to collect system snapshots during early boot stages \cite{latzo2021bringing}.  
Intel’s System Management Mode (SMM) has been used for secure and isolated introspection of memory and CPU state \cite{Reina12}.  
Additional approaches, including coupling SMM with PCI devices, demonstrated hardware-assisted acquisition of register and memory contents \cite{Wang2011}.

Other efforts have targeted specialized environments.  
Acquisition techniques for programmable logic controllers (PLCs) \cite{RAIS2021301196,ZUBAIR2022301336,AWAD2023301513} and baseboard management controllers (BMCs) have been developed.  
BMCLeech, for instance, leveraged BMC capabilities to capture host state without detection \cite{latzo2020bmcleech}.  
Hardware-based methods \cite{BAUER2016S65, 6657268, bulygin2008chipset} and kernel-level remapping of page tables \cite{stuttgen2015acquisition} further illustrate the need for visibility beneath the OS.

However, these works primarily aim at post-mortem forensics or isolated data extraction.  
They do not provide continuous monitoring, runtime integrity guarantees, or coordinated detection across multiple systems. 

\subsection{Positioning of Peacock}

In contrast to static analysis tools, which examine firmware images offline, Peacock captures firmware behavior as it occurs during the boot process.  
Similarly, while prior forensic acquisition frameworks collect snapshots of system state, they do not provide mechanisms for ongoing monitoring or verifying runtime behavior within the UEFI environment.

Peacock addresses this gap by coupling integrity-assured logging with post-boot remote attestation.  
The framework records detailed UEFI Boot and Runtime Service activity during the DXE phase and binds these measurements to hardware-backed evidence.  
Once the OS starts, the recorded measurements are extracted and verified by an external server before being forwarded to a SIEM for rule-based analysis.

Through this design, Peacock contributes a complementary capability to existing firmware security approaches by enabling:

\begin{itemize}[leftmargin=*]
\item runtime telemetry recorded during early boot,
\item integrity protection over collected measurements using attestation evidence,
\item external verification of authenticity and freshness, and
\item integration with enterprise detection systems through structured, rule-based analysis.
\end{itemize}

This combination provides practical visibility into pre-OS activity while supporting scalable detection across multiple systems.
\section{\label{sec:conclusion}Discussion}

This research introduced Peacock, a framework that brings integrity-assured visibility and attested log collection into the UEFI boot process. By combining early-boot telemetry capture, hardware-rooted attestation, and enterprise-scale analysis, the framework addresses a long-standing gap in firmware security: the absence of trustworthy, runtime-level monitoring during UEFI execution.

We implemented the full workflow in virtualized environments and deployed the logging component on physical hardware to validate feasibility across different platform types. The framework was evaluated against four prominent real-world UEFI bootkits: Glupteba, BlackLotus, LoJax, and MosaicRegressor. Across all cases, Peacock successfully supported detection of key behaviors, including unauthorized service modification, Secure Boot abuse, filesystem-based persistence, and coordinated multi-component implants. These results demonstrate that bringing structured and attested firmware telemetry into a central analysis server enables practical identification of threats that have historically operated without meaningful visibility.

While Peacock does not eliminate the underlying classes of firmware vulnerabilities, it raises the difficulty for adversaries by reducing the stealth advantages traditionally associated with pre-OS malware. Attackers must now evade a system that records and attests to boot-phase activity, increasing their operational complexity and reducing their margin for undetected persistence.

This work demonstrates that extending security monitoring into the firmware layer is both feasible and valuable.

\subsection{Enterprise-Scale Detection Advantages}
The Central Analysis System's integration with existing SIEM infrastructure enables detection capabilities that extend beyond isolated firmware log analysis.
By correlating UEFI telemetry with OS-level security data from EDR systems, firewalls, and other security tools, organizations can identify multi-stage attacks that span firmware and operating system layers.
A bootkit that evades firmware-level detection rules may still be identified when its post-boot payload triggers OS-level alerts, with the SIEM correlating these events to reveal the complete attack chain.

Beyond per-system analysis, aggregating UEFI logs across multiple machines enables detection of attacks that may appear benign in isolation.  
A single compromised device may exhibit behavior that falls within normal variance, but when compared against logs from the wider fleet, its deviations become statistically visible.  
Because it is significantly harder for an attacker to replicate the same evasion strategy across many independent systems, cross-system comparison provides an effective mechanism for identifying a small number of corrupted machines even when their individual logs seem unobjectionable.

\subsection{Analyst Interpretation}
As with any rule-based detection system, several rules may also match legitimate firmware activity. 
For example, the ESP-origin detection rule (Table~\ref{tab:esp_result}) correctly identifies Glupteba’s driver loading from the EFI System Partition, but it also flags the presence of legitimate bootloaders that reside on the same partition. 
This behavior is expected: the rules are designed to surface events that warrant analyst attention rather than to unilaterally label them as malicious.  
Accordingly, matched events should be interpreted within their broader execution context, considering both temporal patterns and correlations with other indicators. 
This reinforces the framework’s role as an integrity-assured telemetry source that supports informed decision-making rather than replacing the analyst’s judgment.

\subsection{Forensics Analysis}
The above presented Peacock UEFI Agent can be used in diverse manners, it can be used for targeted forensics analysis.
By embedding the UEFI Agent directly into the target platform, the platform can capture detailed, fine-grained telemetry during the boot process. This comprehensive visibility enables Peacock to serve not only as a real-time integrity and monitoring solution but also as a robust forensic analysis tool.
Investigators can replay and analyze historical logs to trace malicious modifications, detect stealthy persistence mechanisms, or validate the behavior of firmware components under dynamic execution. This capability transforms Peacock into a valuable asset for both proactive security assurance and post-incident investigation.

\subsection{Limitations}
\label{sec:limitations}

While Peacock demonstrates effective detection of diverse firmware threats, several limitations constrain its current capabilities and deployment scope.

\subsubsection{Deployment Requirements}
The framework requires correct integration into UEFI firmware before providing monitoring capabilities.
The assumption that the UEFI Agent loads as the first DXE driver is critical for comprehensive monitoring, and load order manipulation could enable evasion.
Despite these limitations, Peacock provides novel detection capabilities for firmware-level threats that are largely invisible to traditional security solutions.

\subsubsection{Security Boundary}
Despite its comprehensive visibility into the UEFI boot process, the proposed logging approach is constrained by several inherent limitations. First, the system operates within the same execution layer as a potential attacker. Since our UEFI Agent and malicious firmware code both reside in the same environment, a sufficiently privileged adversary could attempt to tamper with the agent’s instrumentation logic, or intercept its hooks.
This co-residency limits the ability to guarantee absolute isolation or full resistance against sophisticated, in-firmware attackers. To mitigate this, we have leverages the TPM as an external root of trust: once the Peacock driver begins execution, each log entry is measured and extended into a PCR.
Any modification to the logging mechanism or to measured components after this point becomes externally visible through mismatched PCR values. While this ensures integrity of logs produced after Peacock initialization and before other DXE drivers execute, it does not protect against modifications occurring before the agent loads, nor does it fully prevent run-time interference by an attacker operating at the same privilege level. Consequently, although TPM-backed measurement provides strong guarantees of log integrity within a defined window of execution, it does not completely eliminate the risks associated with co-layer adversaries in the UEFI firmware environment.

Another limitation arises from the possibility of an operating system–level adversary interfering with or disabling the OS Agent. In such cases, log transmission from the endpoint may be disrupted. One approach to mitigate this risk is to correlate Peacock data with existing EDR solutions: discrepancies between observed endpoint activity and missing UEFI logs can serve as a strong indicator of suspicious behavior. Alternatively, the Peacock UEFI Agent could be extended with a minimal network stack, enabling it to establish a direct connection and transmit logs from within the UEFI environment itself, thereby bypassing reliance on the OS.

\paragraph{\textbf{Hardening Logs Security}}
To further address these limitations, we propose exploring the use of an out-of-band collection mechanism capable of independently gathering UEFI logs, performing integrity measurements, and transmitting the data through a separate network stack. Such a device would introduce an additional security boundary between the attacker and the logging infrastructure, significantly reducing the adversary’s ability to interfere with log generation or transmission and enhancing the overall robustness of the system.

\subsection{Future Research}
Future extensions of the framework may incorporate anomaly detection to identify deviations from typical firmware behavior that are not captured by handcrafted rules. By learning normal patterns of Boot and Runtime Service usage from attested telemetry, such a module could help surface previously unseen or evolving bootkits.
A complementary direction is automated rule generation. Instead of relying solely on analyst-authored signatures, future work could derive rules by correlating recurring behavioral patterns observed in confirmed malicious samples. This would support faster adaptation to new threats and reduce the manual effort required to maintain detection coverage.
These enhancements would shift the system toward a more adaptive, data-driven detection model that complements the current rule-based approach.

\bibliographystyle{IEEEtran}
\bibliography{references}

\clearpage
\appendices
{\Large\textbf {Appendices}}
\section{UEFI Service Tables, Bootkits, and Hooking Techniques}

\subsection{EFI Boot Services Table}
\label{app:gBS}

Key services of \texttt{gBS} include:

\begin{itemize} [itemsep=0.3em,topsep=2pt,leftmargin=*]

    \item Memory Management Services: Functions such as \texttt{AllocatePages} and \texttt{FreePages} manage physical memory allocation during system initialization;
    
    \item Protocol Management Services: Functions like \texttt{InstallProtocolInterface} and \texttt{LocateProtocol} facilitate interaction with UEFI protocols, which abstract hardware and software services;
    
    \item Event and Timer Services: Functions such as \texttt{CreateEvent}, \texttt{WaitForEvent}, and \texttt{SetTimer} enable event-driven programming and asynchronous operation; and 

    \item Image Services: Functions like \texttt{LoadImage} and \texttt{StartImage} handle the loading and execution of UEFI applications and drivers.

\end{itemize} 

\subsection{EFI Runtime Services Table}
\label{app:gRT}

Key runtime services of \texttt{gRT} include:

\begin{itemize}[itemsep=0.3em,topsep=2pt,leftmargin=*]

    \item Variable Services: Functions such as \texttt{GetVariable}, \texttt{SetVariable}, and \texttt{QueryVariableInfo} manage UEFI variables stored in non-volatile memory for secure configuration and data storage;
    
    \item Time Services: Functions such as \texttt{GetTime} and \texttt{SetTime} handle the system clock and real-time timers; and
    
    \item System Reset Services: The \texttt{ResetSystem} function enables controlled platform resets initiated by the firmware.

\end{itemize}

\clearpage

\section{Description of services from the Boot Services Table used throughout the paper}\label{App:Boot Services}

\begin{table}[H]
\centering
%\caption{UEFI Events and Attributes}
\begin{tabular}{|>{\raggedright\arraybackslash}m{3cm}|>{\raggedright\arraybackslash}m{3cm}|>{\raggedright\arraybackslash}m{10cm}|}
\hline
\textbf{Name of Event} & \textbf{Attributes} & \textbf{Description} \\ 
\hline
\multirow{4}{*}{CreateEvent} 
& Type & Specifies the event type, such as `EVT\_TIMER`, `EVT\_RUNTIME`, `EVT\_NOTIFY\_WAIT`, or `EVT\_NOTIFY\_SIGNAL`. \\ 
\cline{2-3}
& Notify Function & Function executed when the event is triggered (optional). \\ 
\cline{2-3}
& Notify Context & Context data passed to the `Notify Function` when the event is triggered. \\ 
\cline{2-3}
& TPL (Task Priority Level) & Priority level for the notification function, such as `TPL\_CALLBACK` or `TPL\_NOTIFY`. \\ 
\hline
\multirow{5}{*}{CreateEventEx} 
& Type & Specifies the event type, as in `CreateEvent` (e.g., `EVT\_TIMER`, `EVT\_RUNTIME`, `EVT\_NOTIFY\_WAIT`, `EVT\_NOTIFY\_SIGNAL`, `EVT\_SIGNAL\_VIRTUA\_ADDRESS\_CHANGE`). \\ 
\cline{2-3}
& Notify Function & Callback function that executes when the event is triggered. \\ 
\cline{2-3}
& Notify Context & Context data passed to the `Notify Function`. \\ 
\cline{2-3}
& TPL (Task Priority Level) & Priority level at which the notify function runs, such as `TPL\_CALLBACK` or `TPL\_NOTIFY`. \\ 
\cline{2-3}
& EventGroup & Unique GUID for an event group, enabling association with system or protocol-related events, such as `gEfiEventExitBootServicesGuid`, `gEfiEventVirtualAddressChangeGuid`, and `gEfiEventReadyToBootGuid`. \\ 
\hline
\multirow{3}{*}{LocateProtocol} 
& Protocol & Pointer to the protocol's GUID (Globally Unique Identifier), identifying the specific protocol to locate. \\ 
\cline{2-3}
& Registration & Registration key returned from `RegisterProtocolNotify` for tracking protocol installations (optional). \\ 
\cline{2-3}
& Interface & Interface instance of the protocol if located successfully. \\ 
\hline
\multirow{3}{*}{LocateDevicePath} 
& Protocol & GUID of the protocol associated with the device path to locate. \\ 
\cline{2-3}
& DevicePath & Pointer to a device path structure that will be updated with the located device path. \\ 
\cline{2-3}
& Device & Pointer to the handle of the located device that matches the specified path. \\ 
\hline
\multirow{2}{*}{HandleProtocol} 
& Handle & Handle on which the protocol interface is to be retrieved. \\ 
\cline{2-3}
& Protocol & GUID of the protocol to retrieve the interface from. \\ 
\cline{2-3}
& Interface & Pointer to the interface instance of the specified protocol. \\ 
\hline
\multirow{4}{*}{OpenProtocol} 
& Handle & Handle on which the protocol is to be opened. \\ 
\cline{2-3}
& Protocol & GUID of the protocol to open. \\ 
\cline{2-3}
& Interface & Pointer to receive the protocol interface. \\ 
\cline{2-3}
& Attributes & Attributes defining how the protocol will be opened (e.g., `EFI\_OPEN\_PROTOCOL\_BY\_HANDLE\_PROTOCOL`, `EFI\_OPEN\_PROTOCOL\_GET\_PROTOCOL`). \\ 
\hline
\multirow{3}{*}{RegisterProtocolNotify} 
& Protocol & GUID of the protocol for which notifications are registered. \\ 
\cline{2-3}
& Event & Event to be signaled each time the specified protocol is installed. \\ 
\cline{2-3}
& Registration & Key used for managing the notification (optional). \\ 
\hline
\end{tabular}
\end{table}

\clearpage

\end{document}